\documentclass[aps,prd,reprint,superscriptaddress,amsmath,amssymb,floatfix]{revtex4-2}
\usepackage{graphicx,bm,booktabs,microtype,hyperref}
\hypersetup{colorlinks=true,linkcolor=blue,citecolor=blue,urlcolor=blue,pdfauthor={Hisham Anwer},pdftitle={Particle Memory and Ra225 Ac225 Heating in Neutron Star Merger Ejecta}}
\newcommand{\HeatingErrorRange}{20 to 27}

\newcommand{\RadialFullError}{0.001}
\newcommand{\RadialFractionError}{0.024}
\newcommand{\ReducedHeatingError}{0.01}
\newcommand{\BenchmarkError}{0.006}

\newcommand{\ChainInstant}{24}
\newcommand{\ChainTenEnergy}{10}
\newcommand{\ChainFirstEnergy}{51}
\newcommand{\ChainEfficiency}{28}
\newcommand{\MixtureEfficiency}{5.5}

\newcommand{\ChainAlphaRatio}{1.284}
\newcommand{\MixtureAlphaRatio}{1.108}
\newcommand{\AlphaRelativeError}{10.8}
\newcommand{\ElectronRelativeError}{3.55}
\newcommand{\AlphaContribution}{2.98}
\newcommand{\ElectronContribution}{2.54}
\begin{document}
\title{Particle Memory and \texorpdfstring{$^{225}$Ra--$^{225}$Ac}{225Ra--225Ac} Heating in Neutron Star Merger Ejecta}
\author{Hisham Anwer}
\email{hisham@cantab.net}
\affiliation{Physics of the Universe Program, Zewail City of Science and Technology, Giza, Egypt}
\affiliation{Department of Physics, Faculty of Science, Cairo University, Giza, Egypt}
\begin{abstract}
Radioactive decay and particle slowing connect nuclear evolution to late heating in neutron star merger ejecta. We introduce a spatial deposition model that retains the emission age of locally trapped particles across dense and dilute ejecta shells. Its analytic treatment yields an incomplete beta kernel and density moment relations for a prescribed loss law. To evaluate heating histories for energy dependent stopping and prescribed nuclear mixtures, we develop a numerical solver. The $^{225}$Ra--$^{225}$Ac sequence links parent decay and daughter ingrowth to delayed alpha deposition. For an initially pure $^{225}$Ra component at the fiducial ejecta parameters, particles emitted before 10 d supply about 24 percent of deposited power at 11 d and 10 percent of deposited energy from 10 to 20 d. Positive cumulative mass quadrature evaluates heating in smooth and broken density profiles while retaining the particle emission history. The resulting heating calculation agrees with the Kasen and Barnes deposition integral to within 0.006 percent and closes the tested particle energy budgets to within $10^{-6}$. This yields a spatially resolved heating prescription that retains the particle energy required for consistent initialization at finite ejecta age.
\end{abstract}
\maketitle

\section{Introduction}
Beta decay of $^{225}$Ra builds the $^{225}$Ac population, whose decay sequence supplies energetic alpha particles to neutron star merger ejecta \cite{Wu2019}. These particles deposit their energy through collisions as the material expands \cite{Barnes2016}. Nuclear ingrowth and particle slowing therefore connect radioactive production to heating at later ages. The deposited power depends on both the surviving nuclei and the energy carried by particles emitted earlier, linking the nuclear decay history to the thermal evolution of the ejecta \cite{Kasen2019}.

Radioactive heating connects neutron star merger nucleosynthesis to kilonova emission \cite{Metzger2020}. Ejecta density, composition and particle confinement determine how efficiently radioactive energy becomes heat. As the ejecta dilute, deposited power samples nuclear emission from progressively earlier ages. Individual alpha decay sequences can then become important \cite{Chen2024}. Nuclear scaling relations connect alpha decay energies and half lives to radioactive heating in merger ejecta \cite{Anwer2026}. Particle memory is the energy retained by particles between their emission and later deposition. The density distribution determines where these particles stop and how earlier nuclear decays contribute to the heating.

Delayed deposition, analytic loss characteristics and density dependent thermalization are established results \cite{Kasen2019,Shenhar2024}. Setzer et al. integrate a local efficiency prescription over a density profile \cite{Setzer2023}. More detailed calculations follow particle degradation and nonlocal propagation \cite{VandenBerg2026,Andalman2026}. These studies establish the importance of delayed deposition, spatial structure and composition.

Particles emitted at the same age encounter different stopping conditions across the ejecta. Connecting radioactive evolution to the resulting heating therefore requires a deposition model that retains both emission age and local density. We introduce an analytic spatial deposition kernel for locally trapped particles in a homologous power law density profile. Retaining emission age during integration over shells couples nuclear evolution to the distinct stopping histories of dense and dilute material. From this kernel, we derive density variance and third moment relations that connect spatially resolved deposition to descriptions with an effective density, power normalization and collision coefficient. Extending the kernel to prescribed radial abundance gradients incorporates spatial variations in the radioactive source. To calculate full heating histories, we develop a numerical solver that follows energy dependent particle slowing and integrates deposition over nuclear emission history. Comparisons across four nuclear ensembles place the resulting heating histories alongside published deposition solutions and a reference implementation. A positive cumulative mass quadrature provides a practical evaluation for structured density profiles. Finite age continuation then gives a direct physical application, with the kernel determining the subsequent heating supplied by particles emitted before the calculation begins.

We first derive particle trajectories under collisional and expansion losses, then integrate their deposition over the ejecta density profile. Coupling the resulting kernel to radioactive decay establishes the heating histories used to examine density averaging and particle memory in calculations begun at finite age.

\section{Particle slowing and expansion}
Expansion and collisions determine the energy $E(t;s,E_0)$ of a particle injected at age $s$ with energy $E_0$. At expansion age $t$, locally trapped particles obey
\begin{equation}
 \dot E=-\frac{x(E)E}{t}-\rho(v,t)c\kappa(E),\qquad E(s;s,E_0)=E_0.
 \label{eq:loss}
\end{equation}
Here $v$ labels a shell, $c$ is the speed of light and $\kappa$ is the stopping coefficient including particle speed divided by $c$. Its units are energy times area per mass. The expansion coefficient is $x=2$ for alpha particles and $x(E)=1+1/\gamma(E)$ for electrons, where $\gamma(E)=1+E/(m_ec^2)$ and $m_e$ is the electron mass.

A power law stopping approximation makes the characteristic analytic. We fix $x$ and take $\kappa(E)/\kappa(E_0)=(E/E_0)^{-g}$ with $g>0$. Numerical fixed power comparisons use $g=0.15$. The energy dependent calculation below evaluates the full stopping functions. With $z=v/v_{\min}$, $R=v_{\max}/v_{\min}$ and density index $n>3$, the profile is
\begin{align}
 \rho(v,t)&=\frac{M}{4\pi v_{\min}^3t^3 I}z^{-n},\nonumber\\
 I&=\frac{1-R^{3-n}}{n-3},\qquad
 \frac{dM}{M}=\frac{z^{2-n}}{I}\,dz.
 \label{eq:rho}
\end{align}
Here $M$ is ejecta mass and $v_{\min},v_{\max}$ bound the shell velocities. Composition is uniform over mass. The fiducial profile has $M=0.023M_\odot$, $v_{\min}=0.1c$, $v_{\max}=0.4c$ and $n=4.5$. The collision time at the inner edge is defined by
\begin{equation}
 t_{c,\mathrm{in}}^2=\frac{\rho(v_{\min},s)s^3c\kappa(E_0)}{E_0},
 \qquad t_c(z)^2=t_{c,\mathrm{in}}^2z^{-n}.
 \label{eq:tc}
\end{equation}
It is constant in homologous expansion. A larger $t_c$ means stronger collisions at the same age and therefore faster stopping.

Homologous dilution separates the injection and receiving ages through $y=s/t$. With $u=E/E_0$ and $\Xi=x(1+g)$, multiplying the equation for $u$ by $(1+g)u^g$ gives a linear equation for $u^{1+g}$ and hence the characteristic \cite{VandenBerg2026}
\begin{align}
 u^{1+g}&=y^\Xi-(1+g)\left(\frac{t_{c,\mathrm{in}}}{t}\right)^2
                  \Psi_\Xi(y)z^{-n},\nonumber\\
 \Psi_\Xi(y)&=\frac{1-y^{\Xi-2}}{\Xi-2},\qquad \Psi_2(y)=-\ln y.
 \label{eq:char}
\end{align}
The continuous limit at $\Xi=2$ is included. A particle stops when the right hand side reaches zero.

Dense inner shells stop an injected cohort first. Their survival boundary in shell velocity coordinates is
\begin{equation}
 z_f(t,s)=\left[\frac{(1+g)t_{c,\mathrm{in}}^2\Psi_\Xi(y)}
                         {t^2y^\Xi}\right]^{1/n}.
 \label{eq:front}
\end{equation}
A cohort of particles born at age $s$ survives in shells $z>z_f$. The boundary moves outward as the cohort ages. Particles remain in their prescribed shells.

\begin{figure}[tbp]
\centering\includegraphics[width=\columnwidth]{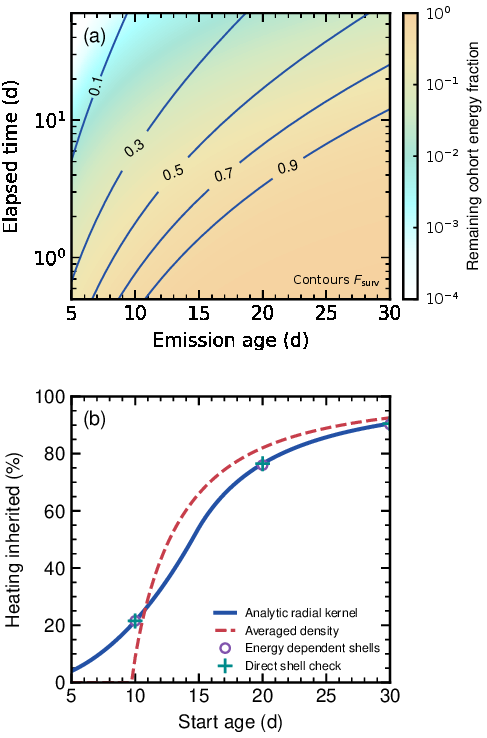}
\caption{Particle survival and deposited heating. The upper color field gives remaining energy of a uniformly injected 6 MeV alpha cohort, divided by initial energy and averaged over mass. Stopped shells contribute zero. Colors are logarithmic, with lower display limit $10^{-4}$. Contours give surviving number fractions 0.1 to 0.9. The lower panel uses fixed $g=0.15$ and $x=2$ for the Ra fed Ac analytic curve and direct shell integration. Separate markers use energy dependent stopping. Their comparison tests the fixed power approximation. Inherited fractions are evaluated one day after each start.}\label{fig:radial}
\end{figure}

For a cohort injected uniformly over mass, the fraction of its particles still above zero energy is
\begin{equation}
 F_{\rm surv}(t,s)=\frac{R^{3-n}-z_b^{3-n}}{R^{3-n}-1},
 \quad z_b=\min\{R,\max(1,z_f)\}.
 \label{eq:mass}
\end{equation}
This is a number fraction. Deposited heating additionally weights each surviving particle by its collisional loss rate. Figure~\ref{fig:radial} separates particle survival, remaining energy and deposited heating.

For alpha particles, the stopping age in a shell follows directly from Eq.~\eqref{eq:char},
\begin{equation}
 t_*(s,z)=s\left[1+\frac{2g}{1+g}
                   \left(\frac{s}{t_c(z)}\right)^2\right]^{1/(2g)}.
 \label{eq:horizon}
\end{equation}
It increases with birth age and with decreasing density. Thus the oldest surviving birth age at a given $t$ is determined uniquely by $t_*(s,z)=t$. In the rapid stopping limit $s\ll t_c(z)$,
\begin{equation}
 t_*(s,z)-s=\frac{s^3}{(1+g)t_c(z)^2}
       \left[1+O\!\left(\frac{s^2}{t_c(z)^2}\right)\right].
 \label{eq:cubic}
\end{equation}
Different shells can occupy different stopping regimes at the same ejecta age.

\section{Spatially integrated deposition}
Dense and dilute shells stop the same injection at different ages. Mass integration retains these histories before convolution with the source.

Particles emitted at the same age form a cohort. For the fixed loss law, a shell deposits power per unit injected energy at rate $K_z(t,s)=t_c(z)^2t^{-3}u^{-g}$ while the cohort survives. Its mass average is
\begin{align}
 \overline K(t,s)&=\frac{t_{c,\mathrm{in}}^2}{t^3I}y^{-xg}\mathcal J(t,s),\nonumber\\
 \mathcal J(t,s)&=\int_{\max(1,z_f)}^R z^{2-2n}
       \left[1-\left(\frac{z_f}{z}\right)^n\right]^{-g/(1+g)}dz.
 \label{eq:Kint}
\end{align}
It vanishes when $z_f\ge R$. The endpoint is integrable because $g/(1+g)<1$. Substitution $w=(z_f/z)^n$ gives
\begin{align}
 \overline K(t,s)&=\frac{t_{c,\mathrm{in}}^2y^{-xg}z_f^{3-2n}}{nt^3I}
                    \mathcal B(t,s),\nonumber\\
 \mathcal B(t,s)&=B_{w_+}(a,b)-B_{w_-}(a,b).
 \label{eq:Kbeta}
\end{align}
Here $a=2-3/n$, $b=1/(1+g)$, $w_-=(z_f/R)^n$, $w_+=\min(1,z_f^n)$ and $B_w(a,b)=\int_0^w\omega^{a-1}(1-\omega)^{b-1}d\omega$ is the incomplete beta function. The limit $z_f\to0$ is finite. This expression integrates local deposition over the density profile before convolving with the source.

Nuclear emission supplies power per mass $q_{{\rm em},j}(s)$ in particle channel $j$. Its deposited contribution is
\begin{equation}
 q_{{\rm dep},j}(t)=\int_0^t q_{{\rm em},j}(s)\overline K_j(t,s)\,ds.
 \label{eq:conv}
\end{equation}
Each channel has its own injection energy and stopping parameters. Nuclear sources follow $\dot{\bm Y}=\bm B\bm Y$, with populations $\bm Y$ per baryon and decay matrix $\bm B$. A parent of rate $\lambda_j$ supplies $q_{{\rm em},j}=\lambda_jY_j\overline E_j/m_u$, where $\overline E_j$ is its channel energy per decay and $m_u$ the atomic mass unit. Branch probabilities enter the matrix and channel energies. Neutrino energy is excluded.

Density variations enter the collision coefficient $C=t_c(z)^2$ through its mass mean $\mu=\langle C\rangle_M$ and variance $V_C=\langle C^2\rangle_M-\mu^2$. The brackets denote a normalized mass integral. Evolution at the mass averaged density uses $C=\mu$. For a recent cohort, $\Delta=t-s\to0^+$, every shell still contains particles. Expansion at fixed receiving age gives
\begin{equation}
 K_C(t,t-\Delta)=\frac{C}{t^3}
 \left[1+g\left(\frac{x}{t}+\frac{C}{t^3}\right)\Delta\right]+O(\Delta^2).
 \label{eq:young}
\end{equation}
Averaging this expression yields the leading discrepancy
\begin{equation}
 \overline K(t,t-\Delta)-K_\mu(t,t-\Delta)
       =\frac{gV_C}{t^6}\Delta+O(\Delta^2).
 \label{eq:variance}
\end{equation}
The variance has units of time to the fourth power. It is an analytic property of the prescribed density profile. Its moments follow directly from Eq.~\eqref{eq:rho},
\begin{equation}
 \langle C^k\rangle_M=t_{c,\mathrm{in}}^{2k}
 \frac{1-R^{3-(k+1)n}}{[(k+1)n-3]I},\qquad k=1,2,3.
 \label{eq:moments}
\end{equation}
The constant and linear terms in Eq.~\eqref{eq:young} distinguish density averaging from loss evolution. Matching the most recent cohort fixes a single coefficient to $\mu$, after which Eq.~\eqref{eq:variance} leaves a nonzero correction for a nonuniform profile.

A free power amplitude does not remove the full history dependence. Consider $\mathcal N_{\rm e}K_{C_{\rm e}}$ at a fixed receiving age. Matching the constant and linear terms requires $C_{\rm e}=\langle C^2\rangle_M/\mu$ and $\mathcal N_{\rm e}=\mu/C_{\rm e}$. The coefficient $C_{\rm e}\ge\mu$ is weighted toward dense material by the collision rate itself. Expansion one order further gives
\begin{align}
 \overline K-\mathcal N_{\rm e}K_{C_{\rm e}}
 &=\frac{g(1+2g)}{2t^9}\mathcal V_3\Delta^2+O(\Delta^3),\nonumber\\
 \mathcal V_3&=\langle C^3\rangle_M-
                  \frac{\langle C^2\rangle_M^2}{\mu}>0.
 \label{eq:thirdmoment}
\end{align}
Both kernels are evaluated at $(t,t-\Delta)$. The strict inequality follows from Cauchy Schwarz for nonuniform positive $C$. The positive quadratic term distinguishes the spatial kernel from the freely normalized single coefficient for a nonuniform profile and $g>0$. The result applies through order $\Delta^2$ to the fixed power loss law before any shell stops. Fits to complete source histories are evaluated separately below.

The excess cohort loss persists until the first shell stops. Write $P=y^\Xi$, $Q=(1+g)\Psi_\Xi(y)/t^2$ and $p=g/(1+g)$. The cohort loss is proportional to $f(C)=C(P-QC)^{-p}$, whose second derivative is nonnegative for $P>QC$. Thus $\langle f(C)\rangle_M\ge f(\mu)$ while every shell survives. Once dense shells stop a cohort, this argument no longer spans the profile. Neither the total history integral nor its normalized inherited fraction has a sign fixed by this inequality. Both must be evaluated with the source retained.

Radioactive material can be concentrated in dense inner shells or dilute outer shells. Let its local emitted power per mass be $q_{\rm em}(s)(I/I_m)z^{-m}$, where $m$ is a prescribed abundance gradient and
\begin{equation}
 I_m=\frac{1-R^{3-n-m}}{n+m-3},\qquad
 dP_m=\frac{z^{2-n-m}}{I_m}dz.
 \label{eq:weightedmass}
\end{equation}
The positive measure $dP_m$ integrates to unity, so $q_{\rm em}$ remains the total emitted power per ejecta mass. The density and stopping medium are unchanged. Positive $m$ concentrates emission inward. For $n+m>3$, the same substitution gives
\begin{align}
 \overline K_m&=\frac{t_{c,\mathrm{in}}^2y^{-xg}z_f^{3-2n-m}}{nt^3I_m}
 [B_{w_+}(a_m,b)-B_{w_-}(a_m,b)],\nonumber\\
 a_m&=2-\frac{3-m}{n}.
 \label{eq:weightedkernel}
\end{align}
The survival boundary and $b,w_\pm$ are unchanged. Channels with different gradients sum linearly. The density moment relations hold with $\langle C^k\rangle_m=\int C^k dP_m$. Source stratification changes the collision coefficient distribution sampled by particles. Uniform composition is recovered at $m=0$. This extension prescribes the source distribution and does not include composition changes to the stopping medium.

The analytic radial kernel agrees with independent shell velocity integration to better than $10^{-6}$ for the tested cohorts, including the continuous limit at $\Xi=2$. Direct integration over birth age checks the source convolution. Independent mass quadrature checks both the linear variance term and the quadratic third moment term. Independent shell integration checks the source weighted kernel for $m=-1,0,1$. Integrated collisional loss, expansion loss and remaining particle energy close the injection energy budget to better than $10^{-6}$ in the tested cohorts.

\section{Radioactive decay and delayed heating}
Production history determines the nuclear populations before particles begin to slow. For initial abundances $Y_{R,0}$ and $Y_{A,0}$ of $^{225}$Ra and $^{225}$Ac, the two member solution is
\begin{align}
 Y_R(t)&=Y_{R,0}e^{-\lambda_Rt},\nonumber\\
 Y_A(t)&=Y_{A,0}e^{-\lambda_At}
 +\frac{\lambda_RY_{R,0}}{\lambda_A-\lambda_R}
 (e^{-\lambda_Rt}-e^{-\lambda_At}).
 \label{eq:ingrowth}
\end{align}
The full tracked chain retains daughter decays and branching. For production at different ages, these populations are integrated over the prescribed production rate. Deposited heating then convolves the resulting positive emission history with the particle kernel. A finite age restart must retain both the nuclei and earlier particles. An initial $^{225}$Ra population defines a Green function component that can be convolved with a production history. A common multiplier of its initial abundance scales all powers and cancels from fractional restart deficits. Relative component abundances still matter in a mixture. The coupled parent and daughter evolution links nuclear ingrowth to delayed alpha deposition. Mixture calculations retain the same particle transport within a broader radioactive source.

For a start age $t_0<t$, the fraction of deposited power supplied by earlier particles in one channel is
\begin{equation}
 D(t,t_0)=\frac{\int_0^{t_0}q_{\rm em}(s)\overline K(t,s)ds}
                    {\int_0^tq_{\rm em}(s)\overline K(t,s)ds}.
 \label{eq:D}
\end{equation}
A common positive multiplier $b(t)$ of the deposited contribution at receiving age $t$ cancels between numerator and denominator. The normalized mean density model therefore retains its original $D$ for every cutoff. Differentiation with respect to $t_0$ recovers $q_{\rm em}(t_0)\overline K(t,t_0)/q_{\rm dep}(t)$. On intervals with nonzero emission, matching $D$ at every cutoff requires the normalized age kernels to agree, a stronger condition than matching total power.

The ages of particles supplying heat can be obtained directly from the same fraction. The deposited power weighted mean lookback is
\begin{equation}
 \overline\Delta_q(t)=\frac{\int_0^t(t-s)q_{\rm em}(s)\overline K(t,s)ds}{q_{\rm dep}(t)}
                   =\int_0^tD(t,\sigma)d\sigma,
 \label{eq:lookback}
\end{equation}
where $\sigma$ is an age cutoff. Interchanging the two positive integrals gives the second equality. This is the age of energy contributing to deposition at $t$, weighted by deposited power.

Nuclear source changes give the fraction a direct dynamical meaning. Apply a common prescribed fractional change $\delta_s\ge-1$ to emission after $t_0$, holding the material and stopping law fixed. Linearity gives
\begin{align}
 q_{\rm em}^{(\delta_s)}(s)&=q_{\rm em}(s)
                [1+\delta_s\mathbf 1_{s\ge t_0}],\nonumber\\
 \frac{q_{\rm dep}^{(\delta_s)}(t)-q_{\rm dep}(t)}{q_{\rm dep}(t)}
                &=\delta_s[1-D(t,t_0)].
 \label{eq:sourcechange}
\end{align}
The indicator $\mathbf 1$ selects the stated interval. The identity holds for total heating under the same change in all channels. Earlier particles spread the heating contribution of the source change over time. An inherited fraction difference $\Delta D$ gives a fractional heating change difference $-\delta_s\Delta D$. Altering collisional stopping changes both the initial rise and the late tail. A change in nuclear decay power can therefore be propagated through the same kernel while retaining energy already carried by older particles.

Rapid stopping and extended storage can coexist across the density profile. The source integrals therefore retain the full particle characteristic. The single energy calculation in Fig.~\ref{fig:radial} uses the Ra fed Ac activity $q_{\rm em}\propto e^{-\lambda_Rs}-e^{-\lambda_As}$, where $\lambda_R=\ln2/(14.9\,{\rm d})$ and $\lambda_A=\ln2/(9.9172\,{\rm d})$.

\section{Radioactive heating in structured ejecta}
Collisions and homologous dilution redistribute radioactive energy across emission ages and ejecta shells. We evaluate deposited power and inherited heating with energy dependent stopping for the specified nuclear ensembles.

Beta heating follows the solar residual abundance pattern and ENDF/B VII.1 decay tables distributed with the HeatingRate reference implementation \cite{HeatingRate}. They span $85\le A\le209$ and $141\le A\le209$, where $A$ is mass number. The first radioactive parent in each tabulated chain has eventual number abundance proportional to the solar pattern. Each selected ensemble is normalized to unit initial mass.

Translead alpha decay is represented by an enriched composition with abundances $(4.0,2.7,4.1,2.7)\times10^{-5}$, respectively, in $^{222}$Rn, $^{223}$Ra, $^{224}$Ra and $^{225}$Ra \cite{Wu2019}. Their total number abundance per baryon is $1.35\times10^{-4}$ and initial mass fraction is 0.03016. This abundance is 13.5 times the approximate total $A=222$--225 upper bound inferred by Wu et al. from AT2017gfo under their heating and luminosity assumptions \cite{Wu2019}. A second case scales the same pattern to $\sum Y_A=10^{-5}$, with mass fraction 0.002234. Both alpha compositions use the solar $85\le A\le209$ beta ensemble for the remaining mass, with its fraction reduced by the initial alpha chain mass fraction. The scaled composition uses the abundance scale of that event dependent bound and is not an event fit.

The tracked alpha paths retain nuclear branching, including both $^{213}$Bi branches for $A=225$. The Ra and Ac half lives use ENSDF and DDEP \cite{ENSDF,DDEP}. The $^{225}$Rn and $^{225}$Fr half lives are 4.66 and 3.95 min, giving a mean delay to $^{225}$Ra of $(4.66+3.95)/\ln2=12.42$ min \cite{ENSDF}. This is short compared with the day scale evolution. These prescribed ensembles isolate beta and alpha heating at fixed nuclear inputs.

Energy dependent collisions use the xenon stopping functions described by Hotokezaka and Nakar \cite{Hotokezaka2020} in both spatial prescriptions. The target mass and charge numbers are 130 and 54. The alpha free electron multiplier is 2 and electron density in the Coulomb logarithm is fixed at $10^5$ cm$^{-3}$. Channels use conditional mean injection energies. Beta power is interpolated in log energy between 33 representative energies from 0.01 to 4.2 MeV. Residual particle energy at 10 keV is deposited promptly. Gamma absorption is also prompt, with $f_\gamma=1-\exp[-(t_\gamma/t)^2]$, $t_\gamma^2=[0.1w+0.003(n-3)/w]\kappa_\gamma M/v_{\min}^2$, $w=v_{\min}/v_{\max}$ and $\kappa_\gamma=0.02$ cm$^2$ g$^{-1}$. This identical global gamma prescription isolates the change in charged particle evolution.

Positive quadrature over cumulative mass retains the density distribution directly. With $\xi\in[0,1]$ denoting mass enclosed above $v_{\min}$,
\begin{equation}
 z(\xi)=[1-\xi(1-R^{3-n})]^{1/(3-n)}.
 \label{eq:massquad}
\end{equation}
We apply Gauss Legendre quadrature in cumulative mass. For a tabulated or piecewise density profile, the shell velocities are obtained by inverting its cumulative mass integral. Positive weights then sum local trajectories while retaining each source history. Eight nodes recover the selected fiducial full heating values to within \ReducedHeatingError\ percent of the 192 node calculation. Inherited fractions, whose integration has a birth age cutoff, require more nodes. The fiducial inherited fraction comparison uses 192 nodes. A second test has $M=0.03M_\odot$, velocities from $0.08c$ to $0.4c$, and a continuous break at $0.18c$. The density indices are 2 inside the break and 6 outside. Beta emission is uniform over mass while alpha abundance varies as $v^{-1}$. The prescribed gamma column is held fixed during radial refinement. With 256 nodes as reference, 8, 32 and 64 nodes give maximum heating errors of 0.098, 0.0105 and 0.0025 percent at the tested ages. The largest inherited fraction difference for eight nodes is 0.028 percentage point. Figure~\ref{fig:benchmark} shows convergence. Splitting the mass integral at the density break gives agreement with the global 256 node rule within 0.0002 percent in power and 0.002 percentage point in inherited fraction at 11 and 21 d. Refining the split rule from 128 to 256 total nodes changes these quantities by less than 0.000004 percent and 0.006 percentage point, respectively. The required node count depends on density and source structure; eight nodes do not guarantee 0.01 percent accuracy for general profiles.

The local channel efficiency of Setzer et al. \cite{Setzer2023}, Appendix A, weights current radioactive emission according to
\begin{equation}
 f_j(t,v)=\frac{\ln(1+X_j)}{X_j},\qquad
 X_j=\frac{2\mathcal A_j}{t\rho(v,t)},
 \label{eq:efficiency}
\end{equation}
with $\mathcal A_\alpha=1.2\times10^{-11}$ and $\mathcal A_\beta=1.3\times10^{-11}$ g cm$^{-3}$ s and $t$ in seconds. The channel power is mass integrated using the same density, radioactive emission and prompt gamma absorption as the trajectory calculation.

The energy dependent solver reproduces a direct evaluation of the single isotope deposition integral of Kasen and Barnes \cite{Kasen2019}, their Eq.~(34), to \BenchmarkError\ percent over $0.1\le t/t_c\le15$. The half life equals the thermalization time, following the convention in their Fig.~3 caption, with $x=2$ and $g=1/2$. Independent clock time integration checks particle trajectories and stiff matrix evolution checks nuclear populations. Refining the energy coordinate, injection ages and deposition quadrature changes selected deposited powers by less than 0.01 percent. Changing radial quadrature from 96 to 192 nodes changes the selected full powers by less than \RadialFullError\ percent and inherited fractions by less than \RadialFractionError\ percentage points. Refining the source change curves from 64 to 128 mass nodes changes the selected fractions by less than 0.02 percentage points. Numerical precision is separate from stopping and nuclear uncertainties.

Density averaging changes the particle history before deposited power is summed. We compare mass integrated local evolution with evolution at $\langle\rho\rangle_M$. Define
\begin{equation}
 \delta_\rho(t)=\frac{q_{\rm mean}(t)-q_{\rm local}(t)}{q_{\rm local}(t)}.
 \label{eq:densityerror}
\end{equation}
Both rates include the entire nuclear emission history and the same gamma absorption. A positive value means that density averaging overestimates the heating. The comparison uses $\langle\rho\rangle_M=\int\rho^2dV/M$. The reference heating prescription uses this mass weighted density to evolve particle losses.

\begin{figure}[tbp]
\centering\includegraphics[width=\columnwidth]{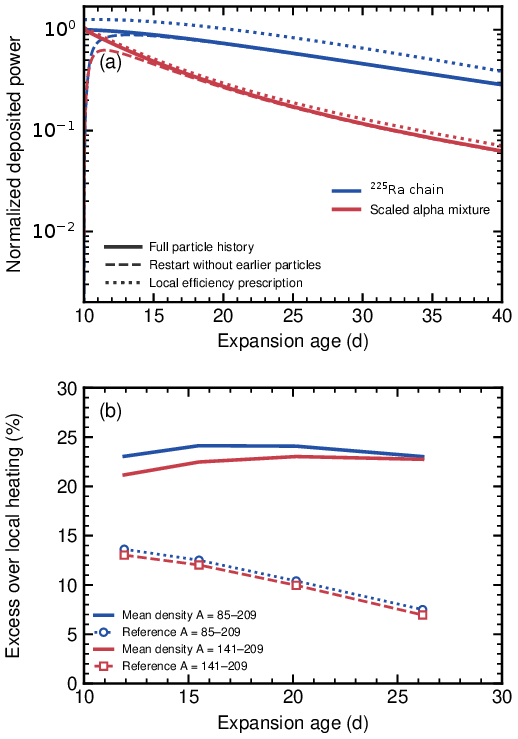}
\caption{Ra chain continuation and reference heating comparison. The chain starts from $^{225}$Ra alone, and the mixture is the scaled alpha ensemble. The upper panel compares full particle histories with removal of particles emitted before 10 d, retaining the same nuclei and future emission. Dotted curves use the published local channel efficiency from Eq.~\eqref{eq:efficiency}. Each source is normalized by its full history deposited power at 10 d. The lower panel gives excess over local energy dependent heating at the four native reference epochs. Solid lines use mass averaged density with matched energy dependent losses. Open markers use the specified HeatingRate implementation \cite{HeatingRate}. Colors distinguish the two solar mass intervals.}\label{fig:density}
\end{figure}

\begin{figure}[tbp]
\centering\includegraphics[width=\columnwidth]{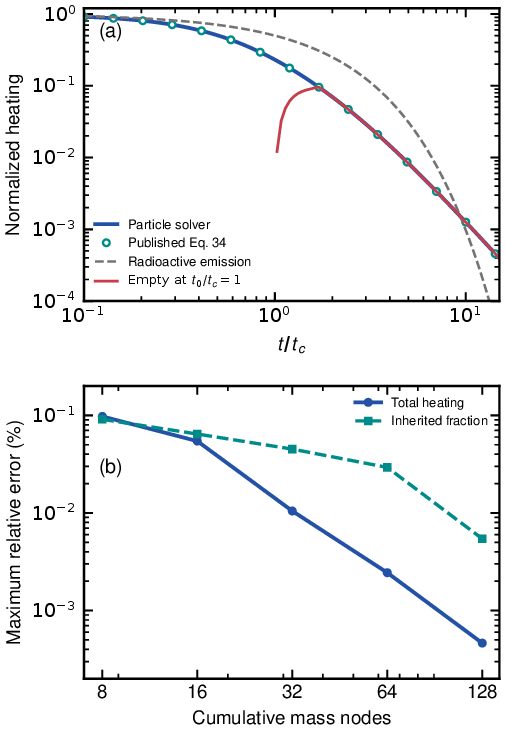}
\caption{Published integral benchmark and structured density convergence. The upper panel evaluates Kasen and Barnes Eq.~(34) for the stated single isotope loss law. The lower panel uses global Gauss Legendre quadrature in cumulative mass for the continuous broken profile and alpha gradient specified in the text. Errors are relative to 256 nodes and maximized over the selected ages. Full heating and inherited fractions require separate convergence checks.}\label{fig:benchmark}
\end{figure}

Evolution at the mass weighted density gives $\delta_\rho>0$ throughout 10 to 30 d in all four fiducial ensembles, with the two solar compositions shown in Fig.~\ref{fig:density}. The excess across all four ensembles is about \HeatingErrorRange\ percent. It persists for both solar beta ensembles and when the total translead abundance is reduced to $10^{-5}$. These full history results apply to the specified profile and loss physics, independently of the recent cohort inequality.

Inner shells stop old cohorts while outer shells retain them. The sign reversal can be isolated by splitting the fixed power integral at $s_{\rm in}(t)$, the oldest birth age surviving at the inner edge. At 20 d in the 6 MeV Ra fed Ac diagnostic, cohorts with $s>s_{\rm in}$ give $q_{\rm local}-q_{\rm mean}=+1.59$ percent of $q_{\rm local}$. Earlier cohorts, already stopped in part of the profile, contribute $-31.64$ percent. Their sum is $-30.05$ percent. The small positive recent cohort correction is therefore outweighed by the changed contribution of partly stopped cohorts. These single energy values explain the sign mechanism separately from the full mixture percentages.

The reference comparison separates two changes of prescription at identical source and receiving ages. Write $q_{\rm ref}$ for the specified HeatingRate implementation \cite{HeatingRate}, $q_{\rm mean}$ for mean density evolution with energy dependent stopping and $q_{\rm local}$ for local evolution. For the two solar ensembles at four epochs from 10 to 30 d, $q_{\rm ref}$ is 6.7 to 12.9 percent below $q_{\rm mean}$, while $q_{\rm mean}$ is 21.2 to 24.2 percent above $q_{\rm local}$. These changes partly offset, leaving the reference 7.0 to 13.6 percent above local heating. Figure~\ref{fig:density} displays both comparisons. The first change includes the fixed power loss law, characteristic convention and native quadrature; the second isolates spatial averaging with matched losses. An independent reconstruction of the specified reference prescription agrees with its outputs within 0.002 percent. This agreement checks implementation, rather than the physical accuracy of either prescribed stopping model.

A normalization change can absorb a nearly constant power offset without restoring the particle history. Multiplying the mean density heating by a constant reduces its maximum residual on the sampled 10 to 30 d interval to 0.9 to 1.3 percent across the four ensembles. This comparison allows the radioactive power itself to change.

Collisional calibration changes stopping while preserving the source. We multiply every charged particle coefficient by $\zeta$, so $t_{c,j}^2\mapsto\zeta t_{c,j}^2$, and leave gamma absorption unchanged. The coefficient minimizes the maximum relative heating error on a half day grid from 10 to 30 d, with $0.08\le\zeta\le2$ and all minima interior. Direct quarter day evaluation checks the result. For collision only calibration, $t_{c,*}/t_c=\sqrt{\zeta_*}$ is 0.69 to 0.71.

Joint calibration also permits a positive output normalization $\mathcal N$. It gives $\zeta=0.91$ to 0.99, $\mathcal N=0.79$ to 0.84 and maximum heating errors of 0.75 to 0.94 percent. This improves only modestly on normalization alone. A common output amplitude cancels from the inherited fraction. At fixed radioactive input it does not specify where the remaining energy resides and is an approximation benchmark rather than a complete energy conserving transport prescription. Holding the solar $85\le A\le209$ joint calibration unchanged gives maximum heating errors of 2.37, 1.76 and 3.34 percent for the other solar, scaled alpha and enriched alpha ensembles, respectively.

The energy carried by earlier particles tests what total power calibration preserves. Table~\ref{tab:comparison} compares all four reductions on both quantities. Joint calibration approximates total heating closely while retaining inherited fraction differences of several percentage points. Equation~\eqref{eq:thirdmoment} establishes an exact kernel limitation in the analytic model, rather than a universal lower bound on the heating residual for a selected source.

\begin{table}[tbp]
\centering\footnotesize\setlength{\tabcolsep}{2.0pt}
\caption{Four reductions compared with local energy dependent deposition across the four ensembles. The power column gives the rounded envelope of signed residuals over 10 to 30 d. The last columns give inherited fraction differences in percentage points one day after the indicated start. A common power normalization preserves those differences exactly.}\label{tab:comparison}
\begin{tabular}{lrrr}\toprule
Reduction & \shortstack{Power\\(percent)} & \shortstack{Start 10 d\\(points)} & \shortstack{Start 20 d\\(points)}\\\midrule
Mass weighted & $+20$ to $+27$ & $-7.4$ to $-5.7$ & $+2.9$ to $+4.0$\\
Power normalized & $\pm1.3$ & $-7.4$ to $-5.7$ & $+2.9$ to $+4.0$\\
Collision calibrated & $\pm8.8$ & $+13.9$ to $+23.6$ & $+9.3$ to $+10.2$\\
Joint calibration & $\pm0.95$ & $-7.0$ to $-2.9$ & $+3.4$ to $+4.2$\\
\bottomrule\end{tabular}
\end{table}

\section{Particle memory at finite age}
Particles emitted before a restart continue to supply heat afterward. The isolated $A=225$ chain starts with $Y_{R,0}=2\times10^{-6}$ per baryon and zero populations in $^{225}$Ac and all other chain members. Its fractional deficits are independent of the initial Ra amplitude. We compare full evolution with a continuation at $t_0=10$ d that retains the same nuclei and future emission while omitting earlier particles. The mixture throughout this section is the scaled alpha ensemble. Figure~\ref{fig:density} compares both sources. For the chain, the instantaneous deficit at 11 d is about \ChainInstant\ percent and the deposited energy deficit integrated from 10 to 20 d is about \ChainTenEnergy\ percent. The deposited energy deficit integrated from 10 to 11 d is about \ChainFirstEnergy\ percent. For the mixture, the deposited energy deficits are about 54 percent from 10 to 11 d and 17 percent from 10 to 20 d. Direct Simpson and trapezoidal time integrals differ by less than 0.1 percentage point in these tests.

These losses arise from omitted particle energy rather than a change in radioactive production. They concern calculations begun at finite age, such as a late heating continuation or a handoff from a nuclear evolution calculation. An evolution that retains the full particle state already includes this contribution. Finite start calculations also carry radiation energy. For example, Bulla’s POSSIS radiation initialization integrates earlier deposited heating with an adiabatic factor to set the starting radiation energy density \cite{Bulla2023}, Eq.~(1). Radiation storage and unthermalized particle energy are distinct state variables. Across the four ensembles, direct local integration gives inherited fractions of 29 to 31 percent at 11 d and 69 to 81 percent at 21 d for starts at 10 and 20 d, respectively.

The empty particle state is a limiting diagnostic. Prompt gamma absorption remains at the start; the removed charged particle contribution builds up from subsequent emission. A practical alternative uses current source power and the local channel efficiency in Eq.~\eqref{eq:efficiency}. This algebraic prescription does not require an empty particle state. Its coefficients encode an approximate slowing history, so its comparison with the energy dependent trajectory includes both that history approximation and differences in stopping physics. At 11 d, the published local efficiency gives about \ChainEfficiency\ percent more chain heating and \MixtureEfficiency\ percent more mixture heating than the trajectory calculation. The alpha efficiency to trajectory ratios are \ChainAlphaRatio\ for the Ra seeded chain and \MixtureAlphaRatio\ for the mixture. Each ratio compares efficiency weighted current alpha power with the history dependent deposition integral in Eq.~\eqref{eq:conv}. Source histories and injection energies enter this comparison even when the algebraic efficiency coefficient is shared. In the mixture, its alpha and electron contributions are \AlphaRelativeError\ and \ElectronRelativeError\ percent above their respective trajectory contributions. Their signed contributions to the total difference are about $+\AlphaContribution$ and $+\ElectronContribution$ percentage points. Prompt gamma deposition is identical. At 11 d, the agreement in total heating is therefore not caused by cancellation between alpha and electron errors. The full particle history provides a common physical reference for assessing both approximations. Figure~\ref{fig:density} distinguishes these approximations. These differences compare complete heating prescriptions under the same specified source and ejecta conditions.

Material dependence enters through both collision strength and profile shape. For the scaled alpha ensemble, varying $M$ from 0.01 to $0.05M_\odot$ at fixed velocities, or multiplying both fiducial velocities by 0.8 to 1.25 at the fiducial mass, gives inherited fractions of 19 to 46 percent and 65 to 87 percent one day after starts at 10 and 20 d. Varying $n$ from 3.5 to 6 at $R=4$, or $R$ from 2 to 6 at $n=4.5$, gives a mean density heating excess of 6 to 30 percent at the tested ages from 10 to 30 d. At fixed $n,R$ and particle injection energy, every coefficient scales as $t_{c,j}^2\propto M/v_{\min}^3$. The source retains its independent decay scales $\lambda_jt_{c,j}$. Changing density therefore cannot collapse different nuclear histories onto a curve of collision age alone. The dependence on $n$ and $R$ remains explicit in the spatial kernel and its moments.

Local trapping and uniform composition define the numerical comparisons. Prescribed abundance gradients extend the analytic kernel. Migration, escape and evolving ionization modify particle transport \cite{Andalman2026}. When photon escape is rapid and radiation storage is small, $L\simeq Mq_{\rm dep}$ carries heating differences into bolometric luminosity. Stored radiation also evolves at intermediate optical depth \cite{Kasen2019,Pognan2025}. Coupling to radiation transport would connect these differences to observations, while stopping measurements would test the collision inputs.

\section{Conclusion}
Spatially resolved particle slowing connects radioactive emission to late heating through the distribution of emission ages. We derive an analytic mass integrated kernel and density moment relations for the prescribed loss law, and develop a numerical solver for the corresponding heating histories. The $^{225}$Ra--$^{225}$Ac sequence links nuclear ingrowth to delayed alpha deposition. For an initially pure $^{225}$Ra component at the fiducial ejecta parameters, omitting particles emitted before 10 d gives about \ChainInstant\ percent less deposited power at 11 d and \ChainTenEnergy\ percent less deposited energy from 10 to 20 d. For the scaled alpha mixture, the published local efficiency gives \MixtureEfficiency\ percent more heating at 11 d, compared with the 29 percent deficit of the empty particle state.

Density moments distinguish total heating from particle history. A free amplitude and one collision coefficient match the leading recent-cohort terms while retaining a positive quadratic difference in the analytic kernel. For the tested source histories, normalization absorbs most of the total power offset, and joint calibration retains inherited fraction differences of several percentage points.

The inherited fraction connects finite age continuation to evolving nuclear sources. Its integral gives the deposited power weighted particle age, and its complement determines how a prescribed emission change reaches the heating. Positive cumulative mass quadrature retains this history across smooth and broken density profiles, with accuracy checked separately for each profile and quantity.

The analytic kernel, density moment relations and numerical solver connect radioactive evolution to particle slowing across the ejecta. Agreement with published deposition benchmarks, independent shell integration and particle energy accounting supports the consistency of the calculation within the prescribed loss physics. Retaining emission age during spatial integration resolves how dense and dilute material convert earlier nuclear emission into later heating. These developments provide a practical treatment of particle memory for evolving radioactive sources and calculations begun at finite age, together with reproducible deposition inputs for subsequent kilonova radiation transport.

\section*{Data and code availability}
The numerical solver, nuclear input tables, benchmark data and scripts required to reproduce the calculations and figures are publicly available in the Zenodo archive Kilonova Particle Memory, version 1.0.2 \cite{ParticleMemorySoftware}.

\end{document}